\documentclass{iopart}

\usepackage{amssymb}

 \usepackage{epsfig}

\def\be{\begin{equation}}
\def\ee{\end{equation}}
\def\bea{\begin{eqnarray}}
\def\eea{\end{eqnarray}}

\def\d{\delta}
\def\g{\gamma}

\def\e{\varepsilon}

\def\l{\lambda}
\def\L{\Lambda}

\def\o{\omega}
\def\O{\Omega_0}
\def\Oo{\Omega}

\def\s{\sigma}

\def\vfi{\varphi}
\def\vr{\vec{r}}

\def\tr{\triangle}

\def\cU{{\cal U}}

\def\pt{{$\cal PT$}}

\newcommand{\Sc}{Schr\"odinger }

\begin{document}

\large

\title{Quadratic pseudosupersymmetry in two-level systems}

\author{Boris F Samsonov and V V Shamshutdinova}
\address{
Department of Physics, Tomsk State
 University, 36 Lenin Avenue, 634050 Tomsk, Russia }

\begin{abstract}
Using the intertwining relation we construct a pseudosuperpartner
for a (non-Hermitian) Dirac-like Hamiltonian describing
 a two-level system interacting
in the rotating wave approximation
 with the electric component of an
  electromagnetic field.
  The two pseudosuperpartners and
  pseudosupersymmetry generators close a quadratic pseudosuperalgebra.
A class of time dependent
electric fields for which the equation of motion for
 a two level system placed in this field can be solved exactly
 is obtained.
 New interesting phenomenon is observed.
There exists such a time-dependent detuning
of the field frequency from the resonance value
 that the probability to populate the excited level ceases to
 oscillate and becomes a monotonically growing function of time
 tending to $3/4$.
 It is shown that near this fixed excitation regime
 the probability
 exhibits two kinds of oscillations.
The oscillations with a small amplitude and a frequency close to
the Rabi frequency (fast oscillations)
 take place at the background of the ones with a
big amplitude and a small frequency (slow oscillations).
During the period of slow oscillations
the
 minimal value of the probability to populate the excited level
 may exceed $1/2$  suggesting for an ensemble of
 such two-level atoms the possibility
 to acquire the inverse population and
 exhibit lasing properties.
\end{abstract}


\vspace{2em}

\textbf{Corresponding Author}: B F Samsonov

\vspace{1em}

E-mail: {\it samsonov@phys.tsu.ru}



\section{Introduction}

The supersymmetry in physics has been introduced in
the quantum field theory for unifying different interactions in
a unique construct \cite{r1}. Supersymmetric formulation of quantum
mechanics is due to the problem of spontaneous supersymmetry breaking
\cite{r2}. Ideas of supersymmetry have been profitably applied to many
nonrelativistic quantum mechanical problems since, and now there are no
doubts that the supersymmetric quantum mechanics
(SUSY QM)
 has its own right to exist
(for recent developments see
a special issue of Journal of Physics A, vol. 34, No 43, 2004).
It is worth noticing that
 most papers in this field deal with the Hermitian Hamiltonians.

Differential equation of Schr\"odinger-like type with a
non-Hermitian Hamiltonian appears in many physical models. One can
cite quantum systems coupled to the environment like a hydrogen
``atom" in an interacting medium subject to a dissipative force
\cite{ND} (see also \cite{Wong}) or different decay or collision
reactions (%
see e. g. \cite{Baz}; for more recent developments see
\cite{RDM}; in \cite{Baye} the method of SUSY QM is involved).
Physical needs initiated a deep mathematical study of
spectral problems with non-Hermitian Hamiltonians in 50th and 60th
of the previous century. The most essential result was first
obtained by Keldysh \cite{Keldysh} who proved the completeness of
the set of eigenfunctions and associated functions
for a regular Sturm-Liouville
problem with a non-Hermitian Hamiltonian. In the books by
Naimark \cite{Naimark} and Marchenko \cite{Marchenko} one can find
good reviews of these studies.

A new impact to studying different properties of non-Hermitian
Hamiltonians is due to the discovery that
 the real character of the spectrum of a non-Hermitian Hamiltonian
  may be in particular related with so-called \pt-symmetry \cite{Bender1}
   and  suggestion to
  generalize the quantum mechanics by accepting non-Hermitian
  Hamiltonians with a real spectrum to describe physical observables
 \cite{Bender2} (see also the review  \cite{Bender3}).
The necessary condition for such a generalization consists in the
possibility to define a Hilbert space with a positive definite
metric which is intimately related with the property of a
Hamiltonian to be diagonalizable (for recent discussions see e.g.
\cite{Bogdan, my1}). This apparently may be assured in many cases
since non-diagonalizable Hamiltonians may be transformed into
diagonalizable ones by SUSY transformations \cite{my2}.
 The latter property
  permits us to suppose that the method of SUSY QM may become an
essential ingredient of the complex quantum mechanics.
This conjecture is also supported by established properties
 of this method not only to offer
  possibility for obtaining new exactly solvable complex
potentials from known ones \cite{Csusy1}
but also to help deeper understanding different properties of complex
potentials \cite{Csusy1,Csusy2}.
In particular, an explicit construction of a superalgebra
involving non-Hermitian Hamiltonians,
 which may be useful in different contexts
i.e. integrability, quantization, different quantum field models
etc, is shown to be possible \cite{susyAlg}
and even is now developed till the notion of pseudosupersymmetry
\cite{Mostafa} and nonlinear pseudosupersymmetry \cite{Pl}.

The relation of the general two-level model described by a
non-Hermitian Hamiltonian acting in the two-dimensional Hilbert
space $\Bbb C^2$ with the pseudosupersymmetry is discussed by
Mostafazadeh \cite{Mostafa}.
In contrast to the approach of this author we
  reduce the time-dependent Schr\"odinger equation for
the two level system,
  interacting
in the rotating wave approximation
 with the electric component of an electromagnetic field,
 with a Hermitian Hamiltonian
 (see e.g.  \cite{books})
   to the one-dimensional stationary Dirac
equation with an effective non-Hermitian Hamiltonian
where the time plays the role of the space variable.
If we considered the spectral properties of the latter Hamiltonian
we would
 define it in the Hilbert space $L^2(0,T)\otimes \Bbb C^2$.
But as we shall see
in our approach
 the spectral parameter
in the Dirac equation is not related with
spectral properties of the two-level system.
Therefore we
will not discuss any spectral  features of this Hamiltonian and
in particular its diagonalizability. Of course, the obtained Dirac
equation is completely equivalent to the initial Schr\"odinger
equation and if one studied it by usual means one would not get
any new information about the two-level system. From this point of
view the method of SUSY QM we are using proves its
extreme efficiency once again.

To find a pseudosuperpartner for the
 given Dirac-like Hamiltonian
we are using
the technique of intertwining operators developed in \cite{Annals}
 for the one-dimensional stationary Dirac equation.
We have to notice that the application of results of
this paper  to our particular problem is not straightforward
since   transformation operators of the general form
do not preserve the very peculiar form of the
effective
Dirac Hamiltonian corresponding to the two-level system.
So, below we show how from the wide variety of possible transformations
one can choose the necessary ones.
In our approach in contrast to \cite{Mostafa}
 the two pseudosuperpartners and
  pseudosupersymmetry generators constructed with the help of
  first order intertwiners
  close a quadratic pseudosuperalgebra.
As it usually happens for the method of intertwining operators
 \cite{BSTMF} if one of the two Hamiltonians is
 exactly solvable the same property takes place for the other.
 In this way
starting from the simplest case corresponding to the famous Rabi
oscillations
 we
 have found new electric fields having time-dependent frequencies for
which the equation of motion of the two-level system has exact solutions.
While analyzing solutions of the \Sc equation we
 have found a new interesting physical phenomenon.
We show that there exists such a time-dependent detuning
of the field frequency from the resonance value
 that the probability to populate the excited level ceases to
 oscillate and becomes a monotonically growing function of time
 tending to $3/4$.
 Of course this is a strictly fixed excitation regime
 similar to resonance.
 We also study how the above probability
 behaves under small deviations from
  this specific regime.
 We have found that when the
 parameters of the model are close enough to the
specific values
 the probability
 exhibits two kinds of oscillations.
The oscillations with a small amplitude and a frequency close to
the Rabi frequency (fast oscillations)
 take place at the background of the ones with a
big amplitude and a small frequency (slow oscillations).
During the period of slow oscillations,
which grows when the parameters of the model approach the above
specific values,
the
 minimal value of the probability to populate the excited level
 may exceed $1/2$ suggesting for an ensemble of
 such two-level atoms the possibility
  to acquire the inverse population and
 exhibit lasing properties.

We have to notice that some of the results we expose below
are known from the previous paper \cite{Bagrov}.
These authors also use a similar intertwining technique but they do
not relate it with the pseudosupersymmetry and do not give any analysis of
solutions this method can provide with.
Moreover,
we give a deeper analysis of restrictions imposed on
transformation operators
 by the features of the two-level system. In particular,
we show that both the new Hamiltonian and solutions of the new
Dirac equation can be expressed in terms of a real-valued
function which is a solution of a second order differential equation
with real coefficients.
Since such equations have real solutions always our analysis opens
the direct possibility to realize chains of transformations
preserving
the form of the Dirac-like Hamiltonian
imposed by the features of the two-level
system.

\section{Preliminary}

The two-level model in the rotating wave approximation
with a possibly time-dependent detuning is described by the
following system of equations (see e.g. \cite{books}):
\be\label{Asist1}
i\dot A_1-fA_1=\xi A_2 \qquad 
i\dot A_2+fA_2=\xi A_1 \ee where $\xi=\frac{1}{2\hbar}E_0d_{12}$,
$d_{12}$ is the matrix element of the dipole interaction operator,
$E_0$ is the amplitude of the electric component of an external
electromagnetic field;
 $f=\frac 12\frac{d}{dt}(\d t)$,
 $\d(t)=\o_{12}-\o(t)$, $\o_{12}=\frac{1}{\hbar}(\e_1-\e_2)$, $\e_1$ and $\e_2$ are energy levels of the
 free atom and $\o(t)$ is the field frequency;
 the dot over the symbol means the derivative with respect to time.
While normalized properly
the functions $|A_1(t)|^2$ and $|A_2(t)|^2$ give occupation
probabilities for the ground and excited states respectively.
 If $\o$ does not depend on time
(hence $f=\frac 12\d=\mbox{const}$) solutions of the system
(\ref{Asist1}) are well-known. For instance,  with the initial
condition $A_2=0$ and $A_1=1$ at $t=0$ we get the well-known
formula \cite{books} for the excited state occupation probability
if  initially  the system is in the ground state
\be\label{C2}
P(t)=
|A_2(t)|^2=\frac{\xi^2}{2\Oo^2}\left[1-\cos(2\Oo t)\right]
\qquad \Oo^2=f^2+\xi^2
\ee
with $2\xi$ known as the Rabi frequency.
The probability (\ref{C2}) is an oscillating function of time (so
called Rabi oscillations). At the resonance ($f=\frac 12 \d =0$) it oscillates
with the Rabi frequency. Therefore the value $\d(t)$ characterizes
the detuning of $\o(t)$ from its resonance value equal $\o_{12}$.
 In Section 5 using the formalism developed
 in Section 4 we shall get time-dependent
functions $f=f(t)$ (and hence $\d(t)$) for which system (\ref{Asist1}) permits
exact solutions.
As we  show below (Section 5) time-dependent
corrections to the detuning that we will
consider
although may change crucially the time-dependent behavior of the
solutions of system (\ref{Asist1}) but they
essentially keep oscillating character
of the probability to populate the excited level
 with the
 frequency close to $2\Omega$.
 Yet, the absence of the  Rabi oscillations
may be considered as oscillations with the same frequency
 but with the zero amplitude since
 they may be obtained as
 corresponding limiting case
 of oscillations with a non-zero amplitude.
So, in our approach the rotating wave approximation is as good as it is
in the classical case
of the electric field of a constant frequency.

Let us rewrite system (\ref{Asist1}) in the matrix form
\be\label{MS}
h_0\Psi=E\Psi \qquad
h_0=\g\partial_t+V_0
\ee
where
\be\label{V0}
V_0= i f_0\s_y
\ee
$\g=i\s_x$, $E=\xi$,
$\Psi= (A_1, A_2)^T$
 (the superscript ``$T$" denotes the transposition)
and we replaced $f$ (which we will call the ``potential") in
(\ref{Asist1})  by $f_0$; $\s_{x,y,z}$ denote the standard Pauli
matrices. Equation (\ref{MS}) is the one-dimensional stationary
Dirac equation with the non-Hermitian Hamiltonian $h_0$ defined by
the potential (\ref{V0}) where $t$ plays the role of the space
variable. By construction the parameters $f_0$ and $E$ are real.
For a fixed value of the dipole momentum of the irradiated system
the parameter
$E=\xi$ is defined by the amplitude of the electric field and,
hence, is not related with spectral properties of the system.
A useful comment is
that since the Hamiltonian of the system (\ref{Asist1}) is
Hermitian, $ H_{sch}=\left(
\begin{array}{cc}f & \xi \\ \xi & -f \end{array}
\right)
$,
the evolution of the two-level system
 is unitary even for a time-dependent function
$f=f(t)$. This means that the $\Bbb C^2$ inner product,
$|A_1(t)|^2+|A_2(t)|^2$, for the Dirac equation (\ref{MS}) is
$t$-independent.

\section{SUSY algebra with non-Hermitian Hamiltonians}

Let us have a non-Hermitian Hamiltonian $h_0$.
We will not consider it as a Hamiltonian acting in a
 Hilbert space but to construct a SUSY
algebra we need adjoint operators which we will introduce in a
formal way. Denote by $h_0^+$ the operator formally adjoint to
$h_0$. As usual the adjoint operation consists in taking the
complex conjugation and transposition, the operator of the first
derivative is skew-Hermitian and $(AB)^+=B^+A^+$.

Let $h_1$ be a ``transformed Hamiltonian" which should be found
together with the transformation operator $L$ by solving the
intertwining relation $Lh_0=h_1L$ and $h_1^+$ be its adjoint. The
later participates in the adjoint intertwining relation
$h_0^+L^+=L^+h_1^+$. It means that the operator $L^+$ transforms
eigenfunctions of $h_1^+$ into eigenfunctions of $h_0^+$.

 Let us suppose that there exists an
operator $J$ such that $h_{0,1}^+=Jh_{0,1}J$
and $J^2=\pm 1$, $J^+=\pm J$
(in general both signs may be accepted).
 Then
from the adjoint intertwining relation it follows that
$JL^+Jh_1=h_0JL^+J$ meaning that the operator $JL^+J$ realizes the
backward transformation from $h_1$ to $h_0$ and the operator $JLJ$
transforms from $h_0^+$ to $h_1^+$.
From here we infer that the
superposition $JL^+JL$ transforms solutions of the equation
(\ref{MS}) into solutions of the same equation meaning that this is
a symmetry operator for this equation.
In the simplest case when $L$ is a differential operator
that we would like to consider this
symmetry operator may be a function of $h_0$,
so we will suppose that $JL^+JL=F_1(h_0)$. By
the same reason the superposition $LJL^+J$ may be a function
of $h_1$ leading to $LJL^+J=F_2(h_1)$.
Moreover, we will also suppose that
$F_2(x)=F_1(x)\equiv F(x)$ is an analytic function.
These properties generalize the known
factorization (polynomial factorization
if $F(x)$ is a polynomial, see e.g.
\cite{Annals,BSTMF})
 properties taking place for the Hermitian case.

 It follows from (\ref{MS}) and (\ref{V0}) that in our case
 $J=\sigma_x$.

Keeping in mind the properties of the operators $L$ and $J$
let us introduce the following matrix operators:
\be
H=\left(
\begin{array}{cc}
h_0 & 0\\ 0 & h_1
\end{array}
\right)  \qquad
Q_1=\left(
\begin{array}{cc}
0 & 0\\ L & 0
\end{array}
\right)  \qquad
Q_2=\left(
\begin{array}{cc}
0 & JL^+J\\ 0 & 0
\end{array}
\right).
\ee
It follows from the intertwining relations that
the operators $Q_1$ and $Q_2$ commute with $H$ and they apparently
are nilpotent.
The above factorization properties are equivalent to the following
anticommutation relation:
$Q_1Q_2+Q_2Q_1=F(H)$.

Now if we identify our $J$ operator with $\eta_-=\eta_+^{-1}$
introduced in \cite{Mostafa}, $J=\eta_-=\eta_+^{-1}$, our $L$
operator with $D$ and $JL^+J$ with $D^{\sharp}$, we conclude that
the operator $Q_2$ becomes pseudoadjoint to $Q_1$, the operators
$H$,$Q_1$ and $Q_2$ close a nonlinear superalgebra and one can
associate a nonlinear pseudosupersymmetry with quantum system
described by the Hamiltonian $H$. In
the next Section we shall show that a quadratic
pseudosupersymmetry may be associated with the two-level system.


\section{Intertwining operators for  two-level Hamiltonians}

To be able to associate a pseudosupersymmetry with the Hamiltonian
given in (\ref{MS}) and (\ref{V0})
we have to find an intertwining
operator and a partner Hamiltonian $h_1$. According to Ref.
\cite{Annals} the intertwining operator $L$ for a matrix equation
such as (\ref{MS}) is defined with the help of a matrix-valued
function $\cU=\cU(t)$ satisfying the equation
\be\label{U}
h_0\cU=\cU \L \qquad \L=\mbox{diag}(\l_1,\l_2)
\ee
called the
``transformation function", as follows:
\be\label{L}
L=\partial_t-W \qquad W=\dot\cU\cU^{-1}\,.
\ee
Here $\l_1$ and $\l_2$ are arbitrary constants.
The operator $L$ transforms a
solution $\Psi$ of equation (\ref{MS}) into a solution $\Phi$ of
the same equation where the matrix $V_0$ is replaced by
\be\label{V1}
V_1=V_0+\Delta V \qquad \Delta V=\g W -W\g\,.
\ee
Here and in the following the subscript $0$ marks
quantities before the
transformation and $1$ marks these after the transformation.
It
is not difficult to see that to preserve the form (\ref{V0}) of
the potential so that $V_1=if_1\s_y$ it is sufficient to take the
transformation function of the form
\be\label{cU}
\cU=\left(
\begin{array}{cc}
u_{11} & u_{11}\\u_{21} & -u_{21}
\end{array}
\right).
\ee
In this case the column-vector
$U_1=(u_{11},u_{21})^T$
is a solution to the initial equation
(\ref{MS}) corresponding to the eigenvalue $\l$ and the
column-vector $U_2=(u_{11},-u_{21})^T$ is a solution to the same
equation with the eigenvalue $-\l$ (note that this symmetry is
built into the system (\ref{MS})!) so that $\L$ in (\ref{U}) has
the form $\L=\mbox{diag}(\l,-\l)$. After some simple algebra one
finds from (\ref{V1}) that  $f_1=f_0+\Delta f$ where
\be\label{Df}
\Delta f=\l \left( \frac{u_{11}}{u_{21}}- \frac{u_{21}}{u_{11}}
\right) -2f_0\,.
\ee

In general, solutions $U_{1,2}$ of equation (\ref{MS})
 from which the matrix $\cU$ is composed,
 $\cU=(U_1,U_2)$,
 are complex, leading to a
 complex-valued potential difference $\Delta f$. For physical
 reasons we require real potentials.
 A necessary condition for $\Delta f$ to be real is
 that the eigenvalue $\l$ be purely imaginary.
Indeed, it is easy to show  that $\l$ cannot be real.
 According to (\ref{Df}) $\Delta f$ is defined by the
expression
$\frac{u_{11\vphantom{I_i}}}{u_{21}}-\frac{u_{21\vphantom{I_i}}}{u_{11}}$.
 Putting
$\frac{u_{11\vphantom{I_i}}}{u_{21}}=\varrho\exp(i\vfi)$
one finds
\be\label{u1121}
\frac{u_{11}}{u_{21}}-\frac{u_{21}}{u_{11}}=
(\varrho-\frac{1}{\varrho})\cos \vfi
+i(\varrho+\frac{1}{\varrho})\sin \vfi
\ee
and our claim follows
from the fact that $\varrho+\frac{1}{\varrho}$ is never equal to
zero. Finally one can prove that $\l^2$ is real
 (cf. \cite{Bagrov}).

Now when the imaginary character of $\l$ is established we see
from (\ref{Df}) that the left hand side of (\ref{u1121}) must be
purely imaginary, which is possible only if $\varrho=1$, meaning
that $u_{11}$ and $u_{21}$ have the same absolute value. Therefore
one can put $u_{11}=\rho\exp(i\vfi_1)$ and
$u_{21}=\rho\exp(i\vfi_2)$. Using the fact that
$U_1=(u_{11},u_{21})^T$ satisfies equation (\ref{MS})
with $E=\l$
and setting
$\l=iR$, where $R$ is real, one gets from (\ref{MS}) a system of
equations for $\rho$, $\vfi_1$ and $\vfi_2$.  Of these equations
we need only
\be\label{vfi}
\dot\vfi_2-\dot\vfi_1-2f_0+2R\sin(\vfi_2-\vfi_1)=0\,.
\ee
If $R=0$
(\ref{vfi}) can be readily  integrated. Suppose  $R\ne 0$.
The
change of the dependent variable in equation (\ref{vfi}),
$\vfi_2-\vfi_1=2\arctan q$, yields for $q$ the Riccati equation
\be
\dot q+2Rq-f_0(1+q^2)=0\,.
\ee
If $f_0=0$ the equation for $q$ is
readily integrated: $q=\exp(-2Rt)$.
 Considering $f_0\ne 0$ one can linearize
(13) by putting $q=-{\dot u}/({uf_0})$, so $u$ is a solution to
the second order equation
\be\label{qE}
\ddot u+(2R- {\dot f_0}/{f_0})\dot u+f_0^2u=0\,.
\ee
Introducing the new variable
$\psi$ by putting $u=\exp(-Rt)\sqrt {f_0}\psi$ one eliminates the
first derivative term from (\ref{qE}) thus obtaining
\be\label{psieq}
\ddot\psi +\left[ f_0^2+\frac 12
\frac{d^2}{dt^2}{\ln}f_0-
\left(\frac 12\frac{d}{dt}\ln f_0-R\right)^2 \right]\psi=0\,.
\ee
This equation has two linearly independent real solutions and,
hence, $\psi$ is defined up to one real constant.
Once $\psi$ is
fixed one calculates $q$:
\be\label{q}
 q=\frac{R}{f_0}-\frac{\dot
f_0}{2f^2_0}-\frac{\dot\psi}{f_0 \psi}
\ee
and the potential
difference $\Delta f=2R\sin(\vfi_2-\vfi_1)-2f_0$:
\be \Delta
f=\frac{4Rq}{1+q^2}-2f_0\,.
\ee

Solution $\Phi$ of the equation $h_1\Phi=E\Phi$ with
$h_1=\g\partial_t+V_1$, $V_1=V_0+\Delta V$, $\Delta V=i\Delta f
\s_y$ can be found by applying the transformation operator
(\ref{L}) to solution $\Psi$ of the equation (\ref{MS}), $\Phi=L\Psi$.
It is easy to see that the matrix $W$ is diagonal
\be\label{w12}
W=\mbox{diag}\left(w_1,w_2\right)\qquad
w_1=-if_0+Ru_{21}/u_{11} \qquad w_2=w_1^*\,.
\ee
and the ratio of the components of the spinor $U_1$ defining $w_1$
 in (\ref{w12}) is also expressible in terms of the function $q$:
\be\label{fracu}
\frac{u_{21}}{u_{11}}=\frac{(1+iq)^2}{1+q^2}\,.
\ee

Finally skipping calculational details but noticing that just in
the same way as it was done in \cite{Annals} one can find the
following factorizations:
\be
JL^+JL=h_0^2-\l^2\,,\quad LJL^+J=h_1^2-\l^2
\ee
with $J=\sigma_1$.
This means that the function $F$ from Section 3 is
$F(x)=x^2-\l^2$,
 the operators $H$, $Q_1$ and $Q_2$ close the quadratic
 superalgebra and the quadratic pseudosupersymmetry underlies the
 two-level system interacting with the electric component of an electromagnetic field.


\section{Application: SUSY transformations of the Rabi oscillations}

In this Section
we show a new physical phenomenon we observed while analyzing
solutions of the system (\ref{Asist1}) obtained using the above
developed technique.

We start with
$\d_0=2f_0=\o_{12}-\o_0 =constant$
(this corresponds to the Rabi oscillations (\ref{C2}))
to get a time-dependent ``potential"
$f_1(t)=f_0+\Delta f(t)=\frac 12 \frac d{dt}[\d_1(t) t]$.
Once $f_1(t)$ is found we calculate the detuning
$\d_1(t)=\o_{12}-\o_1(t)$ by
integrating the previous equation
\be\label{deltat}
\d_1(t)=\frac 2t\int_0^tf_1(t)dt\,.
\ee
We have found that relatively small but time-dependent perturbations of the
field frequency $\o_1(t)$ from its resonance value equal
$\o_{12}$ may influence essentially the
time behavior of the
probability $P_{1}(t)$
 to populate
the excited state level with respect to the constant frequency case.

 If $f_0=\mbox{const}$  equation (\ref{psieq}) for $\psi$  reduces to
\be\label{psidd}
\ddot\psi+\varpi^2\psi=0 \qquad \varpi^2 = f_0^2-R^2=\mbox{const}\,.
\ee
Solutions of this equation
have different properties depending on whether the value
$\varpi^2$ is positive, negative or zero.
We have found that the oscillating behavior of the probability
$P_1(t)$ disappears when $\varpi=0$.
In this case the general solution to equation
(\ref{psidd}) is a linear function of time
$\psi=At+B$ which according to (\ref{q}) gives the following
time dependence of the function $q$:
$q(t)=1-A/(Atf_0+Bf_0)$.
Once $q(t)$ is found one calculates the
``potential difference" with the help of formula (\ref{Df})
 and finally the new ``potential"  $f=f_1(t)$:
\begin{equation}
f_1(t)=f_0-\frac{2A^2f_0}{2 A^2f_0^2t^2-
2 Af_0(A-2Bf_0)t+A^2-2ABf_0+2B^2f_0^2}\,.
\end{equation}
Another restriction leading to the desirable result is
$A=2Bf_0$ which reduces the previous equation to a simpler form
\begin{equation}\label{f1t}
f_1(t)=f_0- \frac{4f_0}{1+4f_0^2t^2}\,.
\end{equation}

Since solutions
$A_{10}(t)$ and $A_{20}(t)$
 of the system (\ref{Asist1}) for
$f=f_0=\mbox{const}$ are known one can find solutions
$A_{11}(t)$ and $A_{21}(t)$
of the same system with $f=f_1(t)$
 by applying the transformation
operator $L$ defined by formulas
(\ref{L}), (\ref{w12}) and (\ref{fracu}) to the previous solution.
In this way imposing the initial condition
  $A_{11}(0)=1$ and $A_{21}(0)=0$  one
finds the probability $P_1(t)$ to populate the excited level at
the time moment $t$ if at $t=0$ only the ground state level is
populated
\bea\nonumber
P_1(t)=|A_{21}(t)|^2=
\frac{\xi^2}{\O^6\left(1+4f_0^2t^2\right)}
\left[16f_0^4\O^2 t^2\cos^2\O t+
\right.
\\  \label{Aoscil}
\left.
4f_0^2\O t\left(\xi^2-3f_0^2\right)\sin 2\O
t+\left(4f_0^2\O^4t^2+\left(\xi^2-
3f_0^2\right)^2\right)\sin^2\O t\right].
\eea
Here $\O=\sqrt{f_0^2+\xi^2}$ and $2\O$ is the frequency of oscillations of
the probability $P_0(t)$ (\ref{C2})
at $f=f_0$.
It is clearly seen that $P_1(t)$ is an oscillating function provided
 $\xi^2\ne 3f_0^2$.
 For  $\xi^2=3f_0^2$ ($\O=2f_0$) the probability becomes
equal
 \be\label{P1t}
P_1(t)=\frac{3f_0^2t^2}{1+4f_0^2t^2}
\ee
which is a function monotonically growing from zero at the initial
time moment till the value $3/4$ at $t\to\infty$.
We have to notice that for a fixed $\xi$
the parameter
$f_0$ is fixed also, $f_0=\xi/\sqrt 3$,
which
by means of formulas (\ref{f1t}) and (\ref{deltat}) fixes
the frequency of the electric field in the unique way.
So,
for the given dipole momentum
this excitation regime is fixed by
 the amplitude of the electric field.
Let us analyze now what is happening with the probability
$P_1(t)$ when the parameters of the model
are close to this exceptional point.

Suppose now  $\varpi^2>0$ and
we will consider it to be close to zero.
In this case the general
solution to equation (\ref{psidd}) may be written as
$\psi=\frac{A}{\varpi}\sin(\varpi t+a+b)$.
The function $q$ as given in (\ref{q}) does not depend on the value
of the coefficient  $\frac{A}{\varpi}$ but we need
this coefficient to
realize the limit $\varpi\to 0$ thus recovering the previously
obtained solution.
Choosing $b$ such that  $\sin 2b=\varpi/f_0$ and $\cos 2b=R/f_0$
but keeping $a$ arbitrary one gets
\be
\frac {\dot \psi}{\psi}=
-\varpi \frac{\varpi-f_0\sin(2\varpi t+2a)}%
{R-f_0\cos(2\varpi t+2a)}\,.
\ee
This leads to the following expression for $q$:
\be
q=\frac{R\cos(2\varpi t+2a)+\varpi \sin(2\varpi t+2a)-f_0}%
{f_0\cos(2\varpi t+2a)-R}
\ee
and finally to the ``potential difference" of the form
\be\label{dfBagr1}
\Delta f(t)=\frac{2\varpi^2}{R\cos(2\varpi t+2a)-f_0}\,.
\ee
This formula has been previously derived by V.G. Bagrov et. al. by other means
\cite{Bagrov}.
Putting $a=\mbox{arctg}\frac{\varpi}{2f_0}-
\frac 12\mbox{arctg}\frac{\varpi}{R}$
one recovers for $f_1(t)=f_0+\Delta f(t)$
the previous result (\ref{f1t}) as the limit
$\varpi\to0$.
 This means that for $\varpi$ close to zero the
probability $P_1(t)$ corresponding to
the potential difference  (\ref{dfBagr1})
should be close to the previous value (\ref{P1t}).
The
analytic expression for $P_1(t)$ is rather complicated and we
will restrict ourselves by
 graphical illustrations.

Let us fix the Rabi frequency
$2\xi$. The function
$\Delta f(t)$ (\ref{dfBagr1}) contains three  parameters
$\varpi$, $f_0$ and $a$. The parameter $f_0$ defines the value
$2\O=2\sqrt{f_0^2+\xi^2}$, which is the frequency of oscillations of
the function $P_0(t)$ given by (\ref{C2}) to which $P_1(t)$ is
reduced when the time dependent correction $\Delta f(t)$ is
absent.
 As it was already mentioned
when $f_0=0$ (resonance case)
the function $P_0(t)$ oscillates with the
Rabi frequency $2\xi$.
The parameter $\varpi$ defines the frequency of the time
dependent correction $\Delta f(t)$ (\ref{dfBagr1})
for $f_1=f_0+\Delta f$
  and the parameter $a$ is responsible for
 the initial value of $f_1(t)$.
The probability $P_1(t)$ is a periodical function
 if $\O$ is commensurable with $\varpi$.
In this case it exhibits two kinds of oscillations, namely,
 fast oscillations
with the frequency $2\O$,
which is close to the Rabi frequency when $f_0$ is close to zero,
taking place at the background of slow
oscillations with the frequency $2\varpi$.


For our numerical illustrations we choose
 $f_0=1$.
 If in standard units  this is
 $1\cdot 10^{11}$ c$^{-1}$
this corresponds to $10^{-11}$ c as the unity of time in our
figures.

Fig. 1a shows the probability $P_1(t)$ for
$\O=2$, $a=0.015$  and $\varpi=1/4$ (solid line)
and $\varpi=1/6$ (dotted line).
 \begin{figure}[th]
\label{fig1}
\renewcommand{\thefigure}%
{\arabic{figure}a}
\begin{minipage}{7cm}
\hspace{4em} \epsfig{file=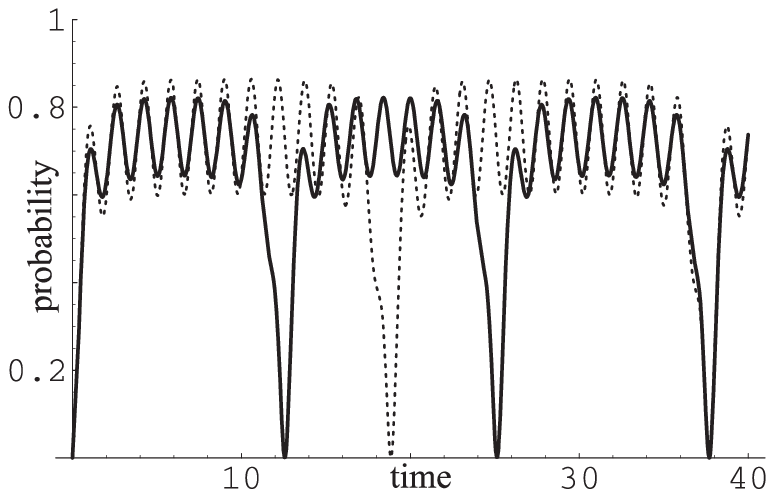, width=6.5cm} \caption{\small
Probability $P_1(t)$ at different values of $\varpi$.}
\end{minipage}
\addtocounter{figure}{-1}
\renewcommand{\thefigure}%
{\arabic{figure}b}
\hspace{1em}
\begin{minipage}{7cm}
\hspace{4em} \epsfig{file=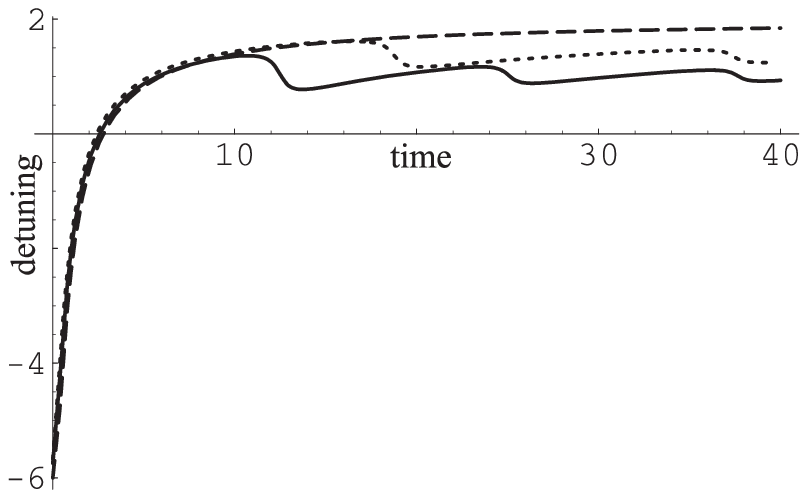, width=6.5cm} \caption{\small
Detuning $\d_1(t)$ at  different values of $\varpi$.}
\end{minipage}
\end{figure}
Fig. 1b illustrates the time behavior of
 the detuning  $\delta_1(t)$ calculated according to
(\ref{deltat}) for  $a=0.015$, $\varpi=1/4$ and $\varpi=1/6$
(solid and dotted lines respectively) together with its limiting value
corresponding to $\varpi=10^{-3}$ and $a=10^{-6}$ (dashed line).
It is clearly seen from  Fig. 1a that the period of slow
oscillations grows when $\varpi$ decreases and fast
oscillations go around the limiting value $0.75$ with the
amplitude increasing with  $\varpi$ decreasing.
 Moreover, Fig. 1b says
that oscillating behavior of $P_1(t)$
is transformed into monotonically growing one
when for $\varpi=0$
the detuning becomes a monotone function of time (dotted line on Fig 1b).
If it acquires some oscillating perturbations the probability
starts to oscillate also.

  \begin{figure}[th]
\label{fig1}
\renewcommand{\thefigure}%
{\arabic{figure}a}
\begin{minipage}{7cm}
\hspace{4em} \epsfig{file=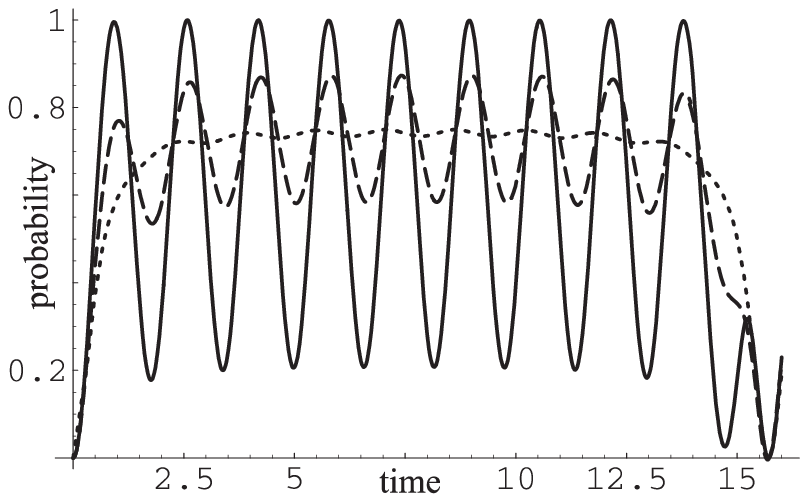, width=6.5cm}
\caption{\small
Probability $P_1(t)$ at different values of $a$.}
\end{minipage}
\addtocounter{figure}{-1}
\renewcommand{\thefigure}%
{\arabic{figure}b}
\begin{minipage}{7cm}
\hspace{3.5em}
\epsfig{file=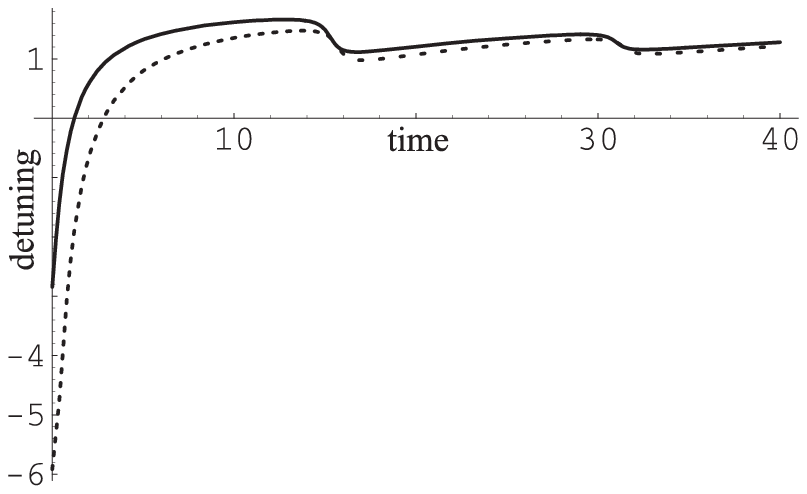, width=6.5cm}
\caption{\small Detuning $\d_1(t)$  at different values of $a$.}
\end{minipage}
\end{figure}
The next two figures show the dependence of the same quantities on
the parameter $a$ which is responsible for the phase shift in
formula (\ref{dfBagr1}) at the fixed value $\varpi=1/5$.
Dotted, dashed and solid lines
 (figure 2a) correspond to
 $a=0$, $a=0.02$ and $a=0.08$ respectively.
 Figure 2b shows
the time dependence of the detuning $\d(t)$ for $a=0$ (dotted line)
and $a=0.08$ (solid line).
From Fig. 2b we can conclude that the parameter $a$  defines mainly
the maximum of the absolute value of the detuning which it takes at $t=0$.
Fig. 2a says that
the amplitude of fast oscillations grows together with $a$.

  \begin{figure}[th]
\label{fig3}
\begin{center}
\begin{minipage}{7cm}
\hspace{4em}
\epsfig{file=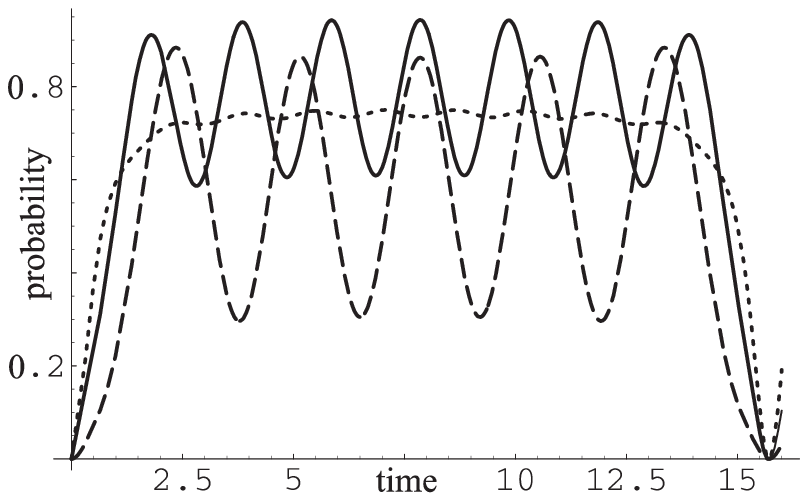, width=6.5cm}
\caption{\small
Probability $P_1(t)$ at different values of $\O$.}
\end{minipage}
\end{center}
\end{figure}
The next figure shows the dependence of $P_1(t)$ from the
frequency of fast oscillations $\O$
at $a=0$ and $\varpi=0.2$.
 Dotted, solid and dashed lines correspond to
 $\O=2$, $\O=1.6$ and $\O=1.2$ respectively.
More it differs from the critical value equal
$2$ corresponding to
 $\xi^2=3f_0^2$, when
the oscillations in
formula (\ref{Aoscil}) disappear, bigger the amplitude
of the fast oscillations becomes.

\section{Conclusion}

Using the technique of intertwining operators for a Dirac-like
system developed in \cite{Annals} we have found time dependent
electric fields for which the equation of
motion for a two-level system placed in this field
obtained after the rotating wave approximation can be solved
exactly.
Pseudosupersymmetry generators constructed with the help of
intertwining operators together with the
super-Hamiltonian close a quadratic deformation of the
superalgebra constructed in \cite{Mostafa}.
 We conclude, hence, that two-level systems
in external electromagnetic fields may have hidden
 quadratic pseudosupersymmetry
 which is responsible for the new phenomenon
 consisting in disappearance of the Rabi oscillations.

\section*{Acknowledgments}

The work is partially supported by the
President Grant of Russia 1743.2003.2
 and the Spanish MCYT and European FEDER grant BFM 2002-03773.
 Authors are grateful to V.G. Bagrov for attracting their attention
 to this problem. BFS is grateful to P. Roy,
 M. Znojil and M. Ioffe for pointing out some useful publications.

\end{document}